\documentclass{aa}

\usepackage{graphicx}
\usepackage{txfonts}
\begin{document}

   \title{Flux rope proxies and fan-spine structures in active region NOAA 11897}

   \author{Y. J. Hou \inst{1,2}, T. Li \inst{1,2}
          \and
          J. Zhang\inst{1,2}
          }

   \institute{Key Laboratory of Solar Activity, National Astronomical Observatories,
   Chinese Academy of Sciences, Beijing 100012, China\\
             \email{yijunhou@nao.cas.cn}
             \and
             University of Chinese Academy of Sciences, Beijing 100049, China
             }

   \date{Received 4 May 2016 / Accepted 28 Jun 2016}


  \abstract
   {Flux ropes are composed of twisted magnetic fields, and connect closely with coronal mass ejections (CMEs).
   Fan-spine magnetic topology is another kind of complex magnetic fields. It has been reported by several authors, and
   is believed to be associated with null-point-type magnetic reconnection.
   }
   {We try to determine the number of flux rope proxies and reveal fan-spine structures in a complex active
   region (AR) NOAA 11897.
   }
   {Employing the high-resolution observations from the Solar Dynamics Observatory (\emph{SDO}) and the Interface
   Region Imaging Spectrograph (\emph{IRIS}), we statistically investigate flux rope proxies in NOAA
   AR 11897 from 14-Nov-2013 to 19-Nov-2013 and display two fan-spine structures in this AR.
   }
   {For the first time, we detect flux rope proxies of NOAA 11897 for total 30 times in 4 different locations during
   this AR's transference from solar east to west on the disk. Moreover, we notice that these flux rope proxies were tracked
   by active or eruptive material of filaments for 12 times while for the rest 18 times they appeared as brightening in the corona.
   These flux rope proxies were either tracked in both lower and higher temperature wavelengths or only detected
   in hot channels. Specially, none of these flux rope proxies was observed to erupt, but just faded away
   gradually. In addition to these flux rope proxies, we firstly detect a secondary fan-spine structure. It
   was covered by dome-shaped magnetic fields which belong to a larger fan-spine topology.
   }
   {These new observations imply that considerable amounts of flux ropes can exist in an AR and the complexity of AR
   magnetic configuration is far beyond our imagination.
   }

   \keywords{Sun: activity -- Sun: atmosphere -- Sun: coronal mass ejections (CMEs) -- Sun: evolution
   -- Sun: filaments, prominences -- Sun: magnetic fields}

   \titlerunning{Flux rope and fan-spine topology}
   \authorrunning{Hou et al.}

   \maketitle
%

\section{Introduction}

A coronal mass ejection (CME) is a large-scale eruption from the solar atmosphere, releasing huge amounts of
mass and magnetic flux into the interplanetary space, and may severely affect the space environment around the
earth (Gosling 1993; Webb et al. 1994). As shown in coronagraph images, a CME generally presents
a three-part structure: the bright front or leading edge, the enclosed dark cavity, and the inner bright
core (Illing \& Hundhausen 1983; Chen 2011). The magnetic flux rope (MFR) is a set of magnetic field lines
winding around a central axis, whose existence could explain the dark cavity's accumulating magnetic energy and
mass in a CME (Gibson et al. 2006; Riley et al. 2008; Wang \& Stenborg 2010). The MFR has been thought
to connect closely with a CME and almost all theoretical models of CMEs require the presence or formation of a
coronal MFR (Forbes 2000). Thus, a complete research on the MFR is necessary to obtain a clear understanding
of CMEs, which will undoubtedly result in accurate forecasts of CMEs and associated space weather.

MFR can be theoretically formed through two ways: bodily flux emergence from below the photosphere into the upper
atmosphere or magnetic reconnection of sheared arcades in the corona. In the emergence model, a twisted MFR is
assumed to exist below the photosphere and then emerge into a pre-existing coronal potential field (Fan 2001, 2010;
Manchester et al. 2004; Magara 2006). Okamoto et al. (2008) found that two opposite polarity regions connected
by vertically weak but horizontally strong magnetic field first grew laterally and then narrowed. And the horizontal
magnetic field changed its orientation accompanying by a significant blueshift. As a result, they suggested that a MFR
emerging from below the photosphere had been observed. It's worth to mention that Vargas Dom{\'{\i}}nguez et al. (2012)
interpreted the same observation as a result of photospheric magnetic cancellation instead of flux emergence.
In the reconnection model, magnetic reconnection between two
bundles of opposite J-shaped loops which have been frequently observed as the sigmoidal structure in the extreme
ultraviolet (EUV) and X-ray lines is able to form the MFR (Canfield et al. 1999; McKenzie \& Canfield 2008;
Liu et al. 2010; Green et al. 2011). And a related possible original mechanism is that the MFR is built up and
heated by the magnetic reconnection in the quasi-separatrix layers (Guo et al. 2013).

Considerable efforts have also been made in numerical simulations to study the formation and dynamic behavior of
MFRs (Forbes \& Priest 1995; Lin et al. 1998; Aulanier et al. 2010). Amari et al. (2000, 2003) simulated
the evolution of a MFR and found that a slow converging motion of the field lines' footpoints toward the polarity
inversion line helps to produce a MFR through magnetic reconnection.
T{\"o}r{\"o}k \& Kliem (2003, 2005) studied the instability of a MFR and found that the kink and/or torus
instability of the MFR can trigger a CME. Olmedo \& Zhang (2010) made a more detailed study and proposed that the
eruption of the MFR can be fully driven by a partial torus instability. Furthermore, setting the observed
photospheric vector magnetic field as the bottom boundary, nonlinear force-free field extrapolation method was
applied to reconstruct the topology of MFR in the corona (Canou et al. 2009; Guo et al. 2010; Jiang et al. 2013).

Direct observations of MFRs have been reported with the help of multi-wavelength observations
from the Atmospheric Imaging Assembly (AIA; Lemen et al. 2012) onboard the Solar Dynamics Observatory (\emph{SDO};
Pesnell et al. 2012). MFRs are usually detected in higher temperature passbands, such as
94 {\AA} and 131 {\AA} (Zhang et al. 2012; Cheng et al. 2013). Furthermore, MFRs sometimes appear one
by one in the same region with a similar morphology, which are named as homologous flux ropes (Li \& Zhang 2013b).
Li \& Zhang (2013a) and Yang et al. (2014) reported that the fine-scale structures of MFRs can be tracked
by erupting material once in a while and could be observed in both higher and lower temperature channels.

Despite a large amount of research on the flux rope, the statistical investigation is rare. Recently,
Zhang et al. (2015) have counted the number of the flux rope proxies over the visible solar disk from
2013 January to 2013 December and pointed out that in some particularly active regions, many flux rope
proxies were detected, which showed some clustering. Our study mainly concerns flux rope proxies in
NOAA 11897 and figures out the locations and number of flux rope proxies during the AR's
evolution from 14-Nov-2013 to 19-Nov-2013. When we observe a set of EUV loops which
collectively wind around a central and axial line, we consider them as a flux rope proxy. Besides the
statistical results, four examples of these flux rope proxies and two fan-spine structures are displayed.
Fan-spine magnetic topology is another kind of complex fields and has been proposed by several authors
(Wang \& Liu 2012; Sun et al. 2013). The fan-spine magnetic topology is believed to play a crucial role
in solar explosive events and direct observations of such a structure has been rare.

The remainder of this paper proceeds as follows. Section 2 contains the observations and data analysis used for our study.
The distribution of the flux rope proxies in NOAA 11897, four typical cases, and two fan-spine structures are presented
in section 3. Finally, we conclude this work and discuss the results in section 4.

\section{Observations and data analysis}

The AIA on board \emph{SDO} observes the multi-layered solar atmosphere, including photosphere, chromosphere,
transition region, and corona in 10 (E)UV passbands with a cadence of 12 s and a spatial sampling of
0.{\arcsec}6 pixel$^{-1}$, among which the 131 {\AA} channel shows flux rope best. Thus we focus on 131 {\AA}
in this study while the 171 {\AA} and 304 {\AA} channels' observations are also presented.
In order to study the relationship between the photospheric magnetic fields of the AR and the flux rope proxies' locations,
line of sight (LOS) magnetograms from \emph{SDO}/Helioseismic and Magnetic Imager (HMI; Schou et al. 2012) with a time cadence
of 45 s and a pixel size of 0.{\arcsec}5 are used as well. Since it's difficult to exclude the possibility that
some polarity was produced by the projection effect when the NOAA AR 11897 was located close to the solar limb,
we adopt the observations from 2013 November 14 to 19 when this AR kept distance away from the limbs.

Moreover, two series of the Interface Region Imaging Spectrograph (\emph{IRIS}; De Pontieu et al. 2014) slit-jaw 1400 {\AA}
images (SJIs) are adopted as well, which are all Level 2 data.
The 1400 {\AA} channel contains emission from the Si IV 1394/1403 {\AA} lines that are formed in lower transition region.
On 2013 November 14, the SJIs in 1400 {\AA} focused on NOAA 11897 were taken from 20:09 UT to 20:35 UT
with a pixel scale of 0.{\arcsec}166, a cadence of 10 s and a field of view (FOV) of 119{\arcsec} $\times$ 119{\arcsec}.
On November 17, the 1400 {\AA} SJIs focused on the same AR were obtained from 12:58 UT to 13:47 UT with a FOV of
167{\arcsec} $\times$ 174{\arcsec}, and the same cadence and pixel size as on November 14.

\section{Results}
\subsection{Overview of the observed flux rope proxies}
\begin{figure*}
\centering
\includegraphics [width=1.\textwidth]{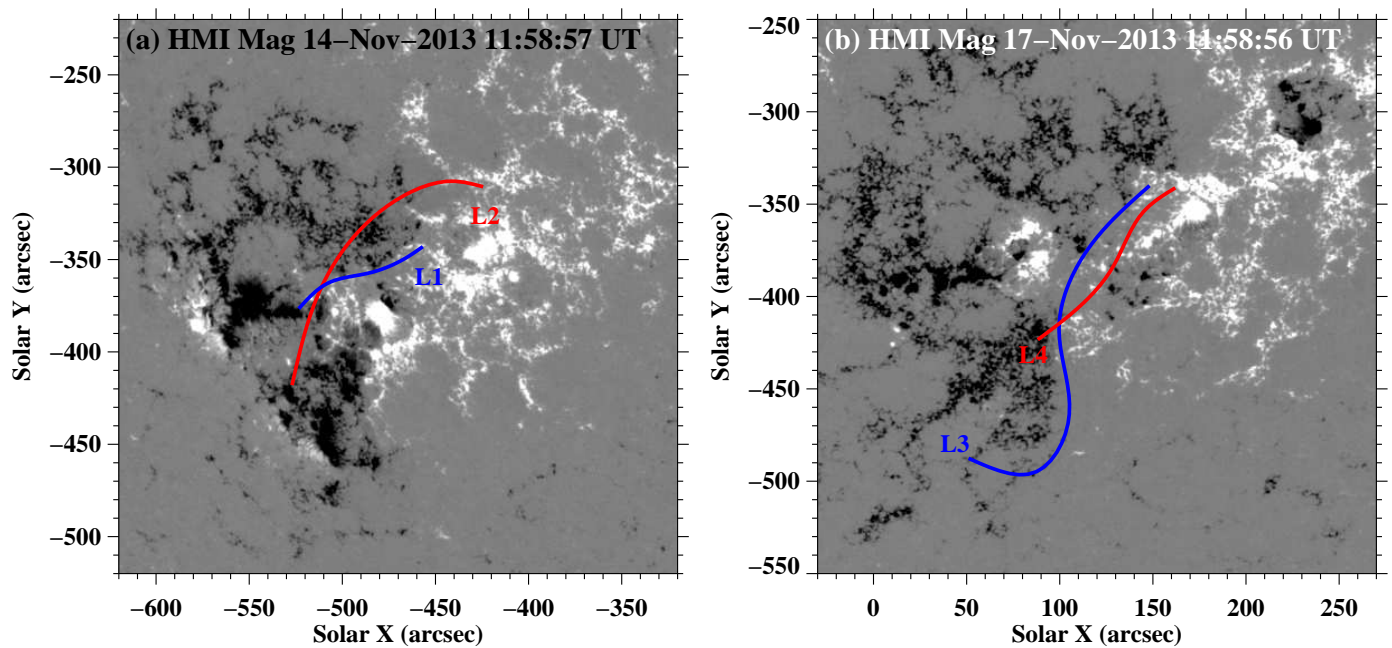}
\caption{
\emph{SDO}/HMI LOS magnetograms showing the magnetic fields of NOAA AR 11897. Red and blue solid curves outline
the 4 locations of detected flux rope proxies in the AR from 2013 November 14 to 19. The red curves mean that
the flux rope proxies on these locations can be detected only in hot EUV passbands, for example, 131 {\AA}, while
the blue curves mean that the proxies on these locations can be detected both in hot and cooler EUV passbands,
such as 131 {\AA}, 171 {\AA}, and 304 {\AA}.
The temporal evolution of the LOS magnetograms is available as a movie (1.mp4) online.
}
\label{fig1}
\end{figure*}

\begin{table*}
\centering
\caption{Distribution of the detected flux rope proxies.}
\centering
\begin{tabular}{c c c c c c c} 
\hline\hline 
& Nov 14 & Nov 15 & Nov 16 & Nov 17 & Nov 18 & Nov 19 \\ 
\hline 
L1 & 3 &  &  &  &  &   \\ \hline
L2 & 8 & 1 &  &  &  &   \\ \hline
L3 &  &  & 2 & 5 & 1 & 1  \\ \hline
L4 &  & 3 & 4 & 1 & 1 &   \\ \hline
\hline 
\end{tabular}
\tablefoot{Flux rope proxies may appear frequently in the same location during several days. The numbers
in the table reveal how many times flux rope proxies appear in the corresponding location on the corresponding day.
}
\end{table*}

During the evolution of NOAA 11897 from 2013 November 14 to 19, we identify flux rope proxies for 30 times in 4 different
locations by using coordinated observations from the \emph{SDO} and the \emph{IRIS}. The locations and general morphologies of these
proxies are shown in Fig. 1 and the daily distribution is summarized in Table 1. The morphologies of some proxies in
one location may change slowly with the evolution
of the underlying magnetic field (see movie 1), but we just classify these flux rope proxies
into the same location by tracing their footpoints' motion. According to our observations, some flux rope proxies can
be detected in all seven EUV channels (304 {\AA}, 171 {\AA}, 193 {\AA}, 211 {\AA}, 335 {\AA}, 94 {\AA}, and
131 {\AA}) that cover the temperatures from 0.05 MK to 20 MK while the others could only be observed in hot
channels such as 94 {\AA} and 131 {\AA}. As a result, we roughly classify these proxies into two types.
To distinguish them, we take the blue and red solid curves to outline these two kinds of flux rope proxies's
locations in Fig. 1, respectively. In the following part, we choose four typical samples to investigate these
flux rope proxies in detail. The first two proxies can be observed in both lower and higher temperature
lines, and the latter two can be only detected in higher temperature lines.

\subsection{Two flux rope proxies detected in both lower and higher temperature wavelengths}
\begin{figure}
\centering
\includegraphics [width=0.49\textwidth]{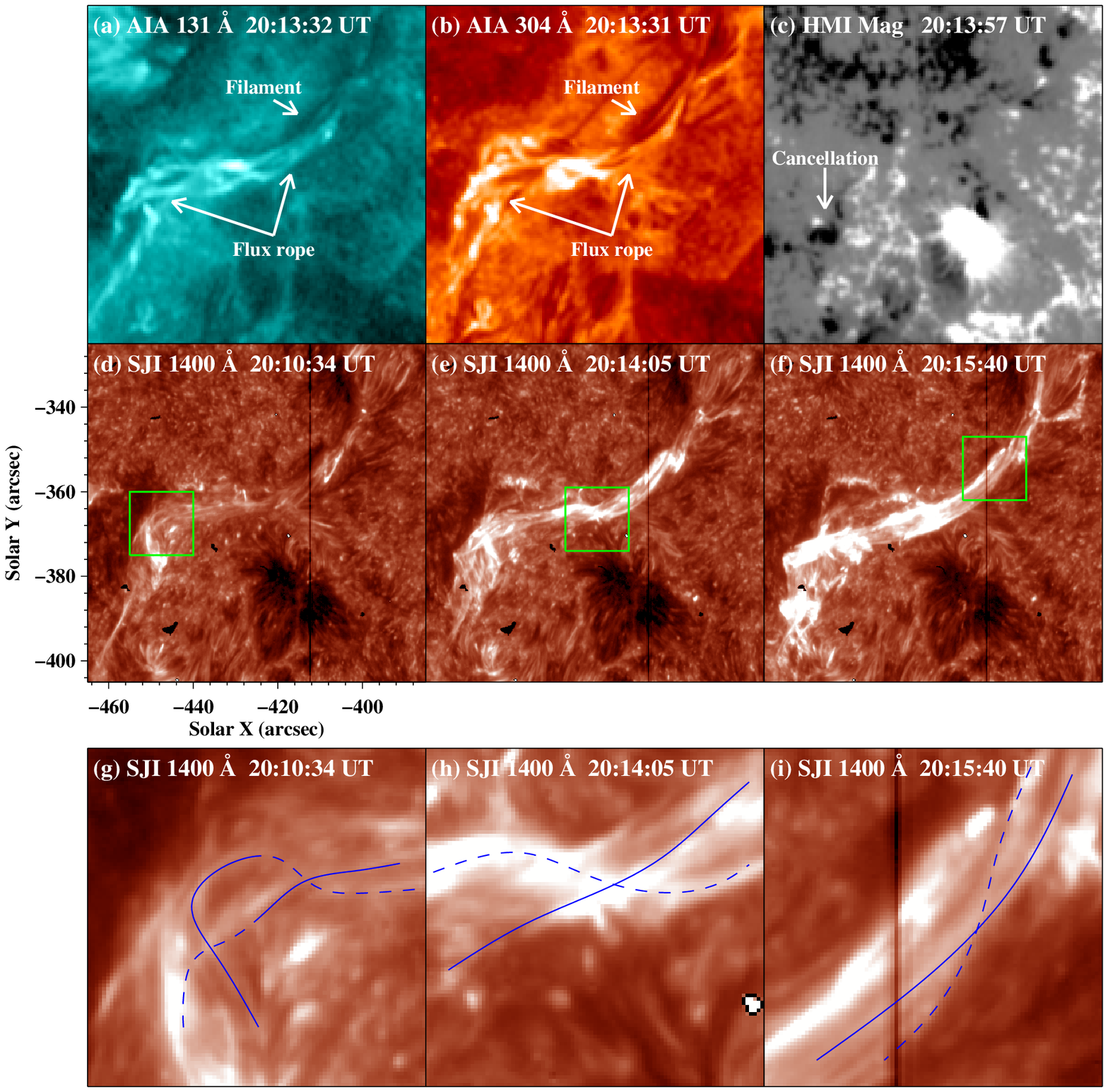}
\caption{
Panels (a)-(c): \emph{SDO} AIA 131 {\AA}, 304 {\AA} images and HMI LOS magnetogram displaying a flux rope proxy on 2013
November 14 in different temperature lines and its underlying magnetic field.
Panels (d)-(f): \emph{IRIS} 1400 {\AA} images showing the brightening of this flux rope proxy.
Panels (g)-(i): three extended 1400 {\AA} images outlined by corresponding green squares in panels (d)-(f).
The blue curves delineate the twisted threads of the flux rope proxy.
The full temporal evolution of the 1400 {\AA}, 131 {\AA}, and 304 {\AA} images is available as a movie (2.mp4)
in the online edition.
}
\label{fig2}
\end{figure}

A flux rope proxy on 2013 November 14 in L1 of Fig. 1(a) was selected and shown in Fig. 2, which was associated with a filament
activation (see movie 2). This filament was located along the neutral line of NOAA 11897 (see panel (c)) and
the rope proxy was brightened by the filament's active material in both lower and higher temperature channels around 20:13 UT
(panels (a)-(b)). In \emph{IRIS} 1400 {\AA} channel, we can observe the brightening of this rope proxy clearly (panels (d)-(f)).
At 20:10 UT, the emission at southeast end of this rope proxy was enhanced, which may result from the underneath magnetic flux
cancellation (see panel (c)). And the twisted structures were observed due to this brightening (see panel (g)). Then the
brightening propagated to the north end and traced the whole flux rope proxy at 20:15 UT (see panels (e)-(f)). Some local
twisted structures were shown in panels (h)-(i). By examining the fine-scale twisted structures in this flux rope proxy,
we roughly estimate that the twist number of this proxy is about 4 $\pi$. From 20:15 UT to 20:18 UT, a rotation motion
of the rope was observed (see movie 2), which can be identified as the unwinding motion of the twisted magnetic field lines
(Yan et al. 2014). Meanwhile, the brightening material moved bi-directionally along the rope proxy. Later, a rotation motion
was detected once again around the southeast end of this rope proxy from 20:20 UT to 20:23 UT.

\begin{figure}
\centering
\includegraphics [width=0.49\textwidth]{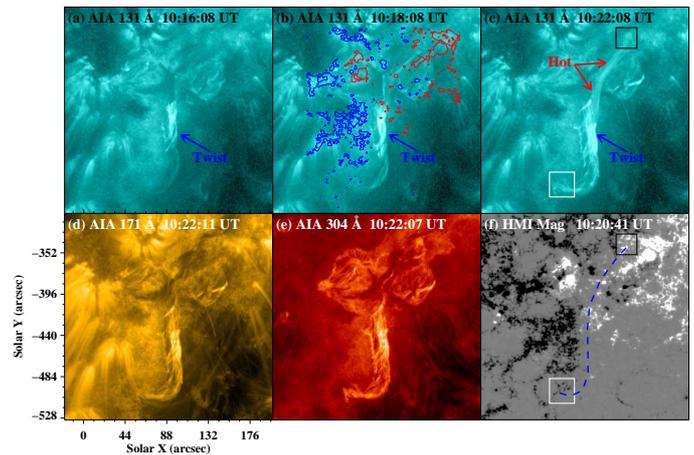}
\caption{
Panels (a)-(c): sequence of AIA 131 {\AA} images showing a flux rope proxy tracked by active material of a
filament on 2013 November 17. The blue and red curves in panel (b) are contours of corresponding negative and
positive magnetic fields, respectively.
Panels (d)-(f): corresponding 171 {\AA}, 304 {\AA} images and HMI LOS magnetogram. In panel (f), the black
square denotes the positive polarity field region where this flux rope proxy's northwest end is rooted while the white
one denotes the negative polarity field region on the other end. These two squares are duplicated to panel (c). The blue
dashed line represents the main axis of this flux rope proxy.
An animation (3.mp4) of the 131 {\AA}, 171 {\AA}, and 304 {\AA} images is available online.
}
\label{fig3}
\end{figure}

Another flux rope proxy that was observed in both lower and higher temperature wavelengths was located at another polarity
inversion line of NOAA 11897 (see L3 in Fig. 1(b)) on 2013 November 17 (see movie 3). Similar to the first
flux rope proxy, the rope proxy shown in Fig. 3 was related to a filament as well. Around 10:16 UT, the filament was initially
disturbed and accompanied by EUV brightening (panel (a)). Then the brightening material moved bi-directionally along the neutral
line (panel (b)) and 10 minutes later, a sigmoid flux rope proxy appeared in both lower and higher temperature wavelengths
(see panels (c)-(e)). Twisted structures (see arrows in panels (a)-(c)) were tracked by the brightening material in both lower
and higher temperature channels. The twist number of this proxy is estimated as 2 $\pi$.
At 10:22 UT, this rope proxy was entirely traced out and the EUV emission of its southeast end was
strengthened after the arrival of brightening materials. After that, the rope proxy's spatial scale decreased gradually as well
as the EUV intensity till the proxy's completely vanishing at 10:38 UT. Nearly the whole of this rope proxy was similar in
different channels except that the north part of the proxy could only be detected in hot channel (see panels (c)-(e)).
Checking the HMI LOS magnetogram, we find that the northwest end of the flux rope proxy was rooted in the positive fields
while the southeast end in negative fields (see panel (f)).

\subsection{Two flux rope proxies detected only in the higher temperature wavelengths}
\begin{figure}
\centering
\includegraphics [width=0.49\textwidth]{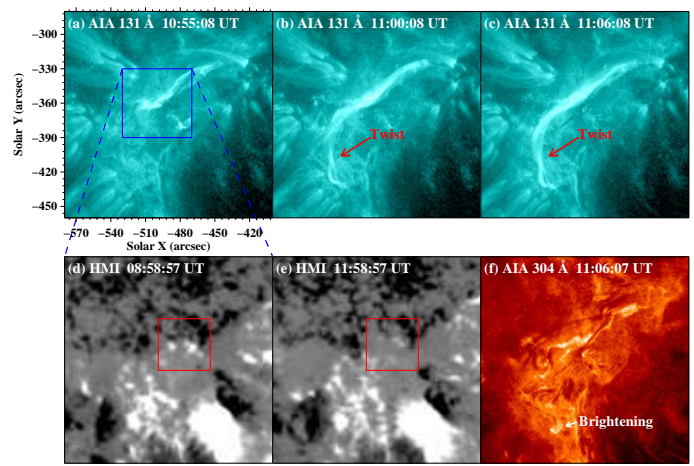}
\caption{
Panels (a)-(c): AIA 131 {\AA} images exhibiting a flux rope proxy on 2013 November 14. The blue square in panel (a)
outlines the FOV of panels (d) and (e).
Panels (d)-(e): corresponding HMI LOS magnetograms showing the evolution of the underneath magnetic fields.
The red squares denote the region where magnetic cancellation occurred.
Panel (f): AIA 304 {\AA} image with the same FOV of panels (a)-(c) displaying this rope proxy in a lower
temperature channel.
The temporal evolution of the 131 {\AA} and 304 {\AA} images is available as a movie (4.mp4) online.
}
\label{fig4}
\end{figure}

Besides the flux rope proxies detected in both lower and higher temperature channels, some rope proxies could be detected
only in hot lines. The first case took place on 2013 November 14 in L2 of Fig. 1(a) and was shown in Fig. 4. In this case,
the flux rope proxy was initiated by a partial eruption of a filament (see movie 4) and the flux cancellation
underlying this proxy was observed as well (panels (d) and (e)). At 10:55 UT, brightening filament material appeared in both
304 {\AA} and 131 {\AA} channels (panel (a)). But during the following ten minutes, the brightening material moved along the
two directions of the filament channel tracking the rope proxy (panel (b)) only in 131 {\AA} passband. And the twisted structures
(see arrows in panels (b)-(c)) were tracked by the brightening material in hot channel as well. Here we estimate that
this proxy's twist number is about 2 $\pi$. The rope proxy was well developed at 11:06 UT (panel (c)) and faded away about 20
minutes later. In the lower temperature line (e.g., 304 {\AA}), we could not observe the main body of this rope proxy except
the brightening footpoints (see arrow in panel (f)).

\begin{figure}
\centering
\includegraphics [width=0.49\textwidth]{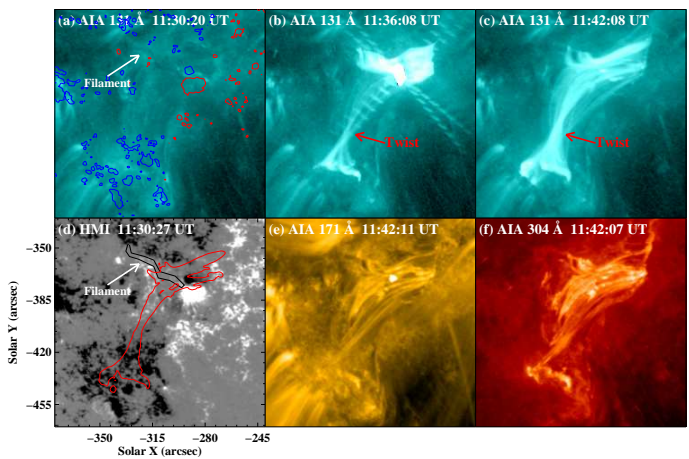}
\caption{
Panels (a)-(c): sequence of AIA 131 {\AA} images displaying a flux rope proxy tracked by a filament's active
material on 2013 November 15. The blue and red curves in panel (a) are contours of corresponding negative and
positive magnetic fields, respectively.
Panels (d)-(f): corresponding HMI LOS magnetogram, AIA 171 {\AA} and AIA 304 {\AA} images. In panel (d), the black
curve outlines the general shape of the filament, and the red curve is the contour of brightness in panel (c).
An animation (5.mp4) of the 131 {\AA},171 {\AA}, and 304 {\AA} channels shown in this figure is available in the
online edition.
}
\label{fig5}
\end{figure}

Another flux rope proxy of this kind was observed on 2013 November 15 in L4 of Fig. 1(b) (see movie 5).
This rope proxy was initiated by a failed eruption of a filament (see Fig. 5(a)). In 131 {\AA} channel, the filament material
began to brighten at 11:32 UT. Then the brightening material moved to the southeast and tracked the rope proxy in the
following ten minutes (panel (b)), which was accompanied by a C7.6 flare. The arrows in panels (b)-(c) point out the twist
of this proxy and we estimate that the twist number is about $\pi$. At 11:42 UT, the rope proxy was well developed
(panel (c)). Then the brightening material moved back to filament again, and the rope proxy faded away ten minutes later.
We outline the general shapes of the proxy and the filament in panel (d), which show the filament's position relative to
the proxy. Similar to the example shown in Fig. 4, this rope proxy's main body cannot be detected in the lower temperature
wavelengths, such as 171 {\AA} and 304 {\AA} (see panels (e) and (f)).

\subsection{Two fan-spine structures}

In NOAA 11897, we also detected two fan-spine structures in addition to those flux rope proxies shown above. The first
structure was observed on 2013 November 18 (see movie 6). At 04:10 UT, a quasi-circular ribbon, whose radius was about 20 Mm,
began to brighten in 131 {\AA}, 171 {\AA}, and 304 {\AA} channels. Subsequently, the bright thread-like structures in 131
{\AA} wavelength appeared and composed a twisted loop bundle with its eastern end connecting to the circular ribbon while the
other end to the remote brightening region (see Figs. 6(a)-(c)). Thus a dome-shaped structure formed and covered the quasi-circular
ribbon. Then the dome and the loop bundle composed an obvious fan-spine configuration.
This event was accompanied by a C2.8 flare. However, the fan-spine configuration could not be observed in the lower temperature
lines, such as 171 {\AA} and 304 {\AA} (see panels (d) and (e)). As a result, we suggest that plasma filled in the fan-spine
structure is pretty hot and exactly above 10 MK (Sun et al. 2013). The circular ribbon corresponded to the intersection of
a fan surface with the photosphere and it was observed in both lower and higher temperature wavelengths during the whole period.
Moreover, an apparent slipping motion of the fine structure along the circular ribbon and the shuffling motion within the hot
loop bundle were clearly observed. At about 04:40 UT, the bright ribbon faded away and the fan-spine configuration disappeared
later at 04:50 UT. Putting sight on the evolution of the underneath magnetic fields (see movie 1), we notice
that a new dipole emerged around 04:30 UT on November 16 (see the orange region in panel (g)) at the location of the circular
ribbon, which was about two days prior to the fan-spine structure's appearance (see the green region in panel (g)). Then the
negative polarity of this new dipole integrated into pre-existing ambient negative fields on the east side while the positive
part kept growing, which was surrounded by the negative fields in the end. Rapid magnetic flux cancellation occurred successively
between the new positive field and the surrounding negative fields. During the cancellation process, the fan-spine structure
brightened again at about 16:00 UT (see the blue region in panel (g)) and a C2.6 flare was detected simultaneously.
By comparing the AIA 131 {\AA} image and HMI LOS magnetogram, we found that the quasi-circular ribbon lay on the surrounding
negative fields and the fan-spine structure connected this ribbon to remote positive fields (see panels (c) and (f)).

\begin{figure}
\centering
\includegraphics [width=0.49\textwidth]{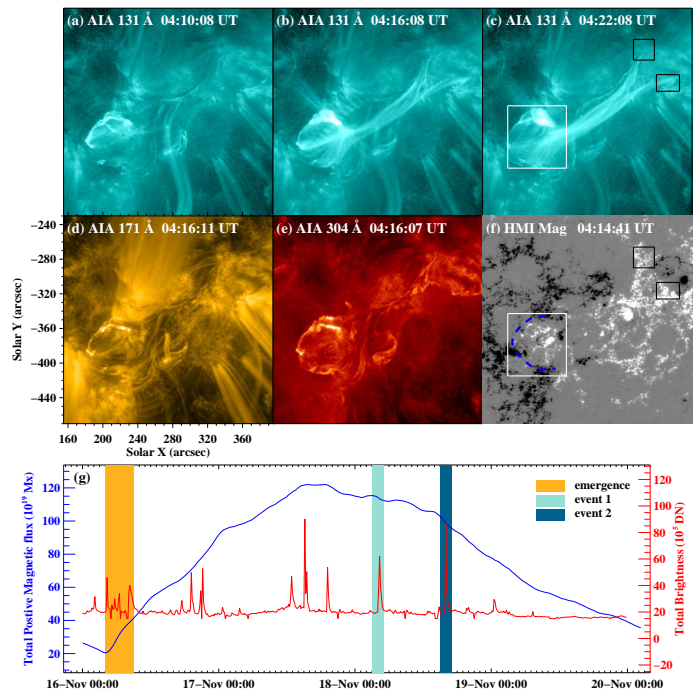}
\caption{
Panels (a)-(c): AIA 131 {\AA} images displaying a fan-spine structure on 2013 November 18.
Panels (d)-(f): corresponding AIA 171 {\AA}, 304 {\AA} images and HMI LOS magnetogram. In panel (f), the black
squares denote the positive magnetic field region where this flux rope proxy's northwest ends are rooted while the white one
denotes the negative field region on the other end and they are duplicated to panel (c).
Panel (g): total positive magnetic flux (blue curve) and brightness (red curve) of the region boxed by white square
in panel (f) from November 16 00:00 UT to November 20 00:00 UT. The orange region marks the start of magnetic field's
emergence while the green and blue regions denote the brightening events of the structure.
An animation (6.mp4) of the 131 {\AA}, 171 {\AA}, and 304 {\AA} images is available online.
}
\label{fig6}
\end{figure}

\begin{figure}
\centering
\includegraphics [width=0.49\textwidth]{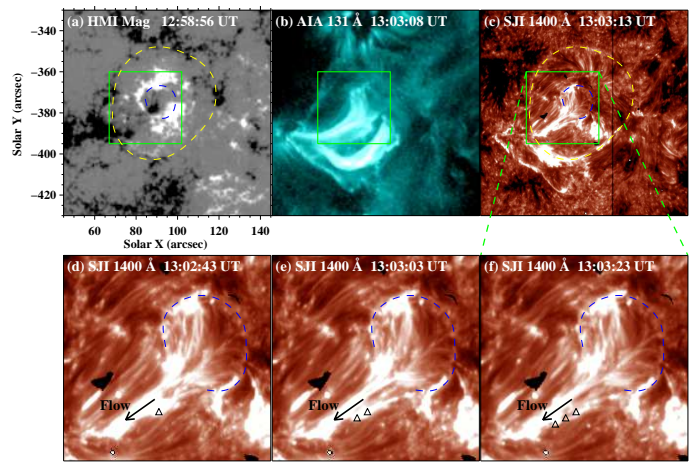}
\caption{
Panels (a)-(c): HMI LOS magnetogram, AIA 131 {\AA} image and SJI of 1400 {\AA} showing the first fan-spine structure
earlier on 2013 November 17. The yellow dashed circles in panels (a) and (c) delineate the bright ribbon while the blue
dashed circles denote the secondary fan-spine structure covered by the dome displayed in Fig. 6. The green squares
outline the FOV of panels (d)-(f).
Panels (d)-(f): extended SJIs of 1400 {\AA} exhibiting the secondary fan-spine structure. The black triangles track
the material's flowing towards southeast.
The full temporal evolution of the 1400 {\AA} and 131 {\AA} images is available as a movie (7.mp4) online.
}
\label{fig7}
\end{figure}

To study the fine structures of this fan-spine structure, we employed the \emph{IRIS} 1400 {\AA} data and compared them
with corresponding \emph{SDO} HMI LOS magnetogram and AIA 131 {\AA} image (see Fig. 7). Although the 1400 {\AA} observations
only covered the period from 12:58 UT to 13:47 UT on November 17, they firstly revealed a secondary fan-spine structure (see movie 7).
The quasi-circle ribbon was clear in 1400 {\AA} channel and we noticed that it was rooted in the surrounding negative fields
(see the yellow dashed lines in panels (a) and (c)). Inside of this ribbon, we detected a smaller quasi-circle ribbon with a
radius of about 6.4 Mm, which lay on the inner ring of positive magnetic fields with negative fields inside (see the blue dashed
lines in panels (a) and (c)). Moreover, a secondary fan-spine configuration above this smaller circular ribbon was observed in
both 131 {\AA} and 1400 {\AA} wavelengths (see the green squares in panels (a)-(c)). In panels (d)-(f), 1400 {\AA} images are
zoomed to show the secondary fan-spine configuration. Similar to the structure shown in Fig. 6, this fan-spine structure connected
the smaller quasi-circle brightening ribbon to remote negative fields near the outer ribbon, which formed a nested dome shape.
During the evolution of this fan-spine structure, a bidirectional flow was observed as well. The flow towards southeast is marked
by tracing brightening plasmoids, whose projected velocities was around 114 km s$^{-1}$ (see the black triangles in panels (d)-(f)).

\begin{figure}
\centering
\includegraphics [width=0.49\textwidth]{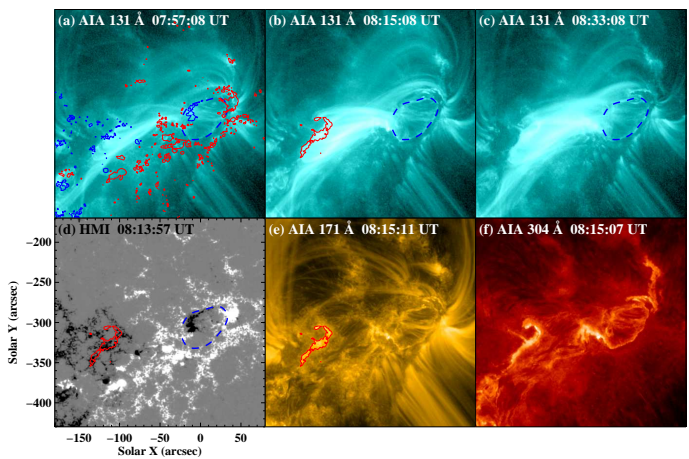}
\caption{
Panels (a)-(c): AIA 131 {\AA} images showing another fan-spine structure on 2013 November 16. The blue dashed circles
outline the quasi-circle ribbon. The blue and red curves in panel (a) are contours of corresponding negative and positive
magnetic fields, respectively.
Panels (d)-(f): corresponding HMI LOS magnetogram, AIA 171 {\AA} and AIA 304 {\AA} images. The red curve in panel (e) is
contour of the remote brightening ribbon and is duplicated to panels (b) and (d).
An animation (8.mp4) of the 131 {\AA},171 {\AA}, and 304 {\AA} channels shown in this figure is available in the online edition.
}
\label{fig8}
\end{figure}

The second fan-spine structure was observed on 2013 November 16 (see movie 8). At about 07:57 UT, a set of loops began to brighten in 131
{\AA} passband, which connected a quasi-circle ribbon to a remote brightening ribbon (see Fig. 8(a)). Then the fan-spine structure was traced
out clearly. Moreover, a sequential brightening of the circular ribbon and an apparent shuffling motion within the hot loop bundle were
detected, which have also been noticed in the first case. But in the lower temperature channels,
we could only see the quasi-circle ribbon and remote brightening ribbon (see panels (e) and (f)). At 08:15 UT, the fan-spine
structure was well developed (see panel (b)). After that, the emission of the loops in 131 {\AA} channel began to weaken. But at about
08:35 UT, the brightening loops appeared in 171 {\AA} and 304 {\AA} channels. We suggest that the brightening loops appearing in
171 {\AA} and 304 {\AA} channels were previous loops in the 131 {\AA} channel which had underwent a cooling process.
Near 09:35 UT, this fan-spine structure disappeared. Investigating the evolution of the underneath magnetic fields (see movie 1),
we find that negative fields emerged near the quasi-circle ribbon's location, which was surrounded by positive fields,
at about 22:00 UT on November 14. It's similar to the case on November 18 that the opposite field's emergence happened several days
earlier than the fan-spine structure's appearance at the quasi-circle ribbon's location. Moreover, the quasi-circular ribbon lay on the
surrounding positive fields and the fan-spine structure connected the ribbon to remote negative fields (see panel (d)).

\section{Conclusions and discussion}
Employing high-resolution observations from the \emph{SDO} and the \emph{IRIS}, we statistically investigate flux rope
proxies in the NOAA 11897 from 2013 November 14 to 2013 November 19. For the first time, we detect flux rope proxies for 30 times
in NOAA 11897 in 4 different locations during 6 days, that is, 5 times per day on average. In this work, we illuminate that flux
rope proxies could appear frequently and be distributed in different locations in an AR. Lites (2005) speculated that flux ropes might
be rather common in normal active regions, which is verified by our observations here. And specially, we exhibit two flux rope
proxies tracked in both lower and higher temperature wavelengths and two brightening rope proxies detected only in hot channels.
For more convincing judgement of these four flux rope proxies, we have roughly estimated their twist numbers. By examining the fine-scale twisted structures of the flux rope proxy in Fig. 2, we estimate that the twist number of this proxy is about 4 $\pi$, i.e., 2 $\pi$ is detected
in Fig. 2(g), $\pi$ in Fig. 2(h), and $\pi$ in Fig. 2(i). As some twisted structures of this proxy cannot be detected, the twist number 4 $\pi$
may be considered as a lower limit. In the late stage of this proxy’s evolution, two rotation motions were detected successively. Combining the
fact that the minimum twist number (4 $\pi$) of this proxy is above the critical value (2.5 $\pi$$\sim$3.5 $\pi$; Hood \& Priest 1981), we suggest
that kink instability possibly took place in this event and the rotation was obviously the unwinding motion of the twisted magnetic field lines.
As a result, the flux rope proxy disappeared in this location (L1) after the rotation motions. For the flux rope proxies in Fig. 3, Fig. 4 and Fig. 5,
no rotation motions took place and their twist numbers have no great changes. So we just estimate an average lower limit of the twist number for
each proxy, which are 2 $\pi$, 2 $\pi$ and $\pi$, respectively. Some authors consider a full turn to be the qualifying property (e.g., Liu et al. 2016),
but others consider a half turn to be sufficient (e.g., Chintzoglou et al. 2015). Our results in this work are consistent with the study of Chintzoglou
et al. (2015) who considered a half turn ($\pi$) to be sufficient for judging a flux rope, as all the four flux rope proxies satisfy such a standard.
In addition to these flux rope proxies, we report on two fan-spine structures in detail.

It is worth mentioning that these 4 locations were not invariant all the time but moved or got deformed slowly with the evolution of
the underlying magnetic fields. Tracing the ends of two flux rope proxies through HMI LOS magnetograms, we decide to classify them into
the same location if these two ropes's both ends are in the same regions, otherwise we determine them as two different locations.
Furthermore, in two locations (L2 and L4 in Fig. 1) of the four ones, flux rope proxies could only be detected in higher temperature
wavelengths (Cheng et al. 2011; Zhang et al. 2012) while flux rope proxies in the rest two locations (L1 and L3 in Fig. 1) have both lower
and higher temperature components (Li \& Zhang 2013). In NOAA 11897, the flux rope proxies which can be seen in both cool and hot lines
(L1 and L3) follow the polarity neutral line much more closely than the proxies which appear only in hot lines (L2 and L4). We suggest that,
in L2 and L4, these flux rope proxies detected only in hot lines may be purely coronal arches at a great height. As for L1 and L3, the proxies
which could be detected in both cool and hot channels may exist at low heights within filament channels. We suggest that, near the polarity
neutral line, cool materials of filament tend to converge within the flux rope structure. Meanwhile, interaction of the opposite polarity fields
around the polarity neutral line may heat up local filament material. As a result, flux rope proxies near the polarity neutral line (L1 and L3)
own complicated temperature feature.

Here we notice that flux rope proxies appeared in one location for several times. It is possible that 7 flux ropes were detected in
4 different locations and repeatedly illuminated for 30 times in total. Since the flux rope proxies in L3 (L4) lasted for 4 days and showed
very different complexity of geometry when they reappeared, we suggest that there were 3 (2) flux rope proxies in L3 (L4). In L3, we consider
the proxy detected twice on Nov 16 as the first one in this location, the proxy detected for 6 times on Nov 17 and Nov 18 as the second one,
and the proxy detected once on Nov 19 as the third one. Moreover, in L4, the proxy detected for 7 times on Nov 15 and Nov 16 is identified as
the first one in this location, and the proxy detected twice on Nov 17 and Nov 18 is identified as the second one. For L1 and L2, we think that
only one flux rope proxy on each of these locations was detected and repeatedly illuminated.
The flux rope proxies which appeared in one location for several times may also be described as homologous flux ropes according to the work
of Li \& Zhang (2013b). They defined the homologous flux ropes as follows: (1) originating from the same region within the same AR;
(2) whose endpoints are anchored at the same location; (3) whose morphologies resemble each other.
Synthesizing Table 1 and Fig. 1, we can clearly find that flux rope proxies appeared with remarkable frequency in L2, L3, and L4.
During the evolution of this AR's magnetic field (see movie attached to Fig. 1), the positive polarity field underlying the L2's
north-western ends moved north-westward along the neutral line slowly while the negative field underlying south-eastern ends
kept generally still. Thus we consider that frequent appearance of flux rope proxies in L2 may result from the underlying
magnetic field's lasting shear motion (Amari et al. 1999, 2000). As for L3, Figure 1(b) shows clearly that it lay above the main
polarity neutral line of NOAA 11897. Considering long-term existence of this main neutral line, it is explicable that flux rope proxies were
observed for 9 times in L3 from November 16 to November 19. In L4, flux rope proxies were detected for 9 times from November 15 to November 18,
which may result from the underlying strong magnetic fields. Flux rope proxies in this location connected a pair of strong magnetic
fields, which existed throughout the entire observation period. Moreover, this pair of strong magnetic fields kept canceling with surrounding
opposite polarity magnetic fields and flux cancellation is thought as the primary formation mechanism of flux ropes (Savcheva et al. 2012a, b).
We follow the evolution of the AR and find that three pairs of magnetic fields emerged. Flux rope proxies were detected for 27 times
in the emerging and stable phase of the magnetic flux. In the decaying phase, only 10$\%$ flux rope proxies (3 times) were detected,
implying an unbalanced distribution of the flux ropes over the entire AR evolution.
Note that none of these flux ropes has been observed to erupt in the end and all of them just faded away gradually.

In-depth research on the fan-spine configuration has been proposed by several authors (Wang \& Liu 2012; Sun et al. 2013). In the work
of Sun et al. (2013), flux emergence was believed to result in a largely closed fan-spine topology. The sequential brightening of
the circular ribbon and the apparent shuffling motion within the hot loop bundle indicated the slipping-type reconnection (Li \& Zhang 2015).
Here we report on two examples which have conspicuous fan-spine configuration, apparent sequential brightening along the circular ribbon and
shuffling motion within the loop bundle in 131 {\AA} passband. And new flux emergence took place several days prior to the fan-spine
structure's appearance in both of the two cases. Priest \& D{\'e}moulin (1995) showed that reconnections may also happen in
quasi-separatrix layers (QSLs) where magnetic connectivity is continuous but with steep gradient. Field lines may continuously
slip within QSLs to exchange their footpoints, which is an extension of the instantaneous break-and-paste scenario (Aulanier et al. 2006, 2007).
The fan-spine topology (Lau \& Finn 1990) often arises with new flux emergence into a pre-existing field. Part of the new field
becomes parasitic and surrounded by the opposite fields. Then a null point forms in the corona (T{\"o}r{\"o}k et al. 2009).
Energized particles from reconnection near the null point flow along the QSLs, brightening their footpoints, which is the
quasi-circler ribbons in our work. For the first time, we detect a secondary fan-spine structure on 2013 November 17. In the spine,
the bidirectional flow was observed, and we suggest that the flow may be caused by the null-point-type reconnection as well.
Moreover, we notice EUV irradiance delays in lower temperature channels. The brightening loops
connecting quasi-circle ribbon to remote brightening ribbon were formed in the fan-spine structure and were
heated to over 10 MK (Sun et al. 2013). The delay of the brightening loops' appearance in cooler channels, such as 171 {\AA} and 304 {\AA},
suggests a cooling process (Liu et al. 2013).

\begin{acknowledgements}
The data are used courtesy of \emph{SDO} and \emph{IRIS} science teams. \emph{SDO} is a mission of NASA's Living With a Star Program.
\emph{IRIS} is a NASA small explorer mission developed and operated by LMSAL with mission operations executed at NASA Ames Research
center and major contributions to downlink communications funded by ESA and the Norwegian Space Centre.
This work is supported by the National Natural Science Foundations of China (11533008 and 11303050) and the Strategic Priority Research
Program$-$The Emergence of Cosmological Structures of the Chinese Academy of Sciences (Grant No. XDB09000000).
\end{acknowledgements}

%
%

\clearpage

\end{document}